\documentclass{sig-alternate}
\usepackage{epsfig}
\begin{document}
\title{Co-Evolution of Friendship and Publishing in Online Blogging Social Networks}
\author{D.~Zinoviev$^\dagger$ and S.~Llewelyn$^+$\\%
$^\dagger$Mathematics and Computer Science Department, Suffolk University\\%
Boston, MA 02114, USA\\%
dzinoviev@suffolk.edu\\%
$^+$Department of Computer Science, Stony Brook University\\%
Stony Brook, NY 11794, USA\\%
sllewelyn@cs.stonybrook.edu%
}
\maketitle

\begin{abstract}
In the past decade, blogging web sites have become more sophisticated and influential than ever. Much of this sophistication and influence follows from their network organization. Blogging social networks (BSNs) allow individual bloggers to form contact lists, subscribe to other blogs, comment on blog posts, declare interests, and participate in collective blogs. Thus, a BSN is a bimodal venue, where users can engage in publishing (post) as well as in social (make friends) activities. In this paper, we study the co-evolution of both activities. We observed a significant positive correlation between blogging and socializing. In addition, we identified a number of user archetypes that correspond to ``mainly bloggers,'' ``mainly socializers,'' etc. We analyzed a BSN at the level of individual posts and changes in contact lists and at the level of trajectories in the friendship-publishing space. Both approaches produced consistent results: the majority of BSN users are passive readers; publishing is the dominant active behavior in a BSN; and social activities complement blogging, rather than compete with it.  
  \end{abstract}

\section{Introduction and Prior Work}
In the past decade, blogging web sites such as
Blogger~\cite{Blood04},
DreamWidth~\cite{Lafayette09},
LiveJournal~\cite{Lafayette09}, and
Tumblr~\cite{marquart10} have become more widespread, more
technologically sophisticated, and more socially influential than
ever. Much of this sophistication and influence stems from their
network organization. Blogging social networks (BSNs) allow individual
bloggers to form contact (``friend'') lists, subscribe to their
friends' blogs, comment on selected blog posts, declare and share
common interests, and participate in communities, or collective
blogs. Thus, a BSN is a socio-semantic network~\cite{roth2010}---a bimodal venue where users engage in publishing
(write blog posts) and social (make friends) activities.

Proper networking aspects of massive online social networks (MOSNs),
including BSNs, have been extensively researched in the past ten
years. MOSN static organization and macroscopic dynamics at the level
of nodes and links: scale-free degree distribution,
shrinking diameter, and densification---have been discussed 
in~\cite{ahn2007,Golbeck07,novak06,leskovec05,pearson2002} and other papers. Analysis
of microscopic behavior at the level of individual member-to-member
messages and message flows can be found
in~\cite{golder2007,leskovec2008,zanette2002}, etc.

Similarly rich literature exists on blogging topics, patterns, and
behaviors~\cite{Gotz2009,Guadagno08,Nardi04,Zhao12}. It covers
information diffusion (including epidemic diffusion), social and
personal motivation for blogging, and bloggers' anthropology.

In a recent study, Both and Cointet~\cite{roth2010} propose that the social and semantic dimensions are co-determined. However, they do not look at the dynamics of individual friendships and posts.

In this paper, we explore the co-evolution of social and publishing
activities in
LiveJournal~\cite{boyd2006,Zakharov07}. LiveJournal was
started in 1999 by American programmer Brad Fitzpatrick and sold to
Russian media company SUP Media in 2007. At the time of writing,
LiveJournal hosts approximately 40 million individual blogging accounts and
communities.

Given the limited amount of time that users spend browsing  social and blogging networks, it is interesting to find out the distribution of networking (social) and publishing (blogging)
activities from the point of view of an individual member of a BSN, as
well as the evolution of this distribution over time.

The rest of the paper is organized as follows: in
Section~2, we explain the data acquisition methods and
present some descriptive statistics of the acquired data;
Sections~3 and~4 contain the microscopic and
macroscopic data analyses, respectively. Section~5
compares the results of both studies. In Section~6, we
conclude.

\section{Data Collection\label{method}}

LiveJournal positions itself as an open blogging platform with a
public application programming interface (API)\footnote{In fact, LiveJournal
provides several APIs, including RSS XML and Atom XML for the most recent posts
and FOAF XML and plain text interface for contact
lists. It is also possible to download profile pages in
HTML and parse them directly.}. The LiveJournal administration
encourages retrieval of publicly available data for the purpose of
research, provided that researchers follow certain guidelines (for
example, applications should make no more than 5 requests per second to the
LiveJournal servers).

We have collected longitudinal publishing and friendship data of approximately 2,000
randomly chosen LiveJournal users over the period of 140 days, starting in August
2011. Sixty-one of these accounts later turned out to be abandoned by their owners, leaving us with 1,836 live accounts.  For each user $U$ and for each day $t_i$, we recorded the
total number of posts over the lifetime of the blog $P^U_i$ and the number of friends $F^U_i$. (We will omit
the upper index whenever it can be inferred from the context.) Thus,
all tuples $\{t_i,P_i,F_i\}$ pertaining to the user $U$ form
a trajectory $T^U$ of the user's blog in the 3-dimensional space
$\{t,P,F\}$.

\begin{figure}[tb!]\centering
\strut\epsfig{file=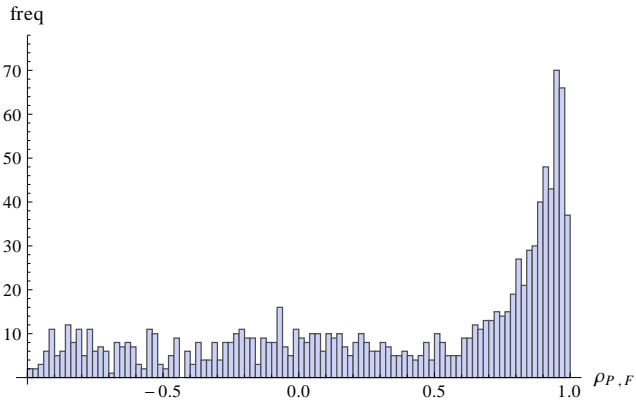,width=\columnwidth}
\caption{\label{Figure1}~Distribution of correlations $\rho_{P,F}$ for all trajectories $T$.}
\end{figure}

The changes of the numbers $P$ and $F$ along a trajectory represent
the publishing and social activities of the user,
respectively:

\begin{align}
p(t)&=P'(t)\big\vert_{t=t_i}\approx\frac{P_{i+1}-P_i}{t_{i+1}-t_i}\\
f(t)&=F'(t)\big\vert_{t=t_i}\approx\frac{F_{i+1}-F_i}{t_{i+1}-t_i},
\end{align}

\noindent%
where $P(t)$ and $F(t)$ are the continuous (interpolated) versions of $P_i$ and $F_i$. In general, $p(t)\ge0$, unless the blogger decides to delete already published posts (a rare but possible condition). There are no restrictions on $f(t)$. We will refer to $P(t)$ or $F(t)$ as $G(t)$ whenever the corresponding formul\ae{} are equally applicable to both of them.

Several trajectories in our collection had unusually large differences between consecutive values ($|G_{i+1}-G_i|\gg1$) or unusually large overall ranges ($G_\text{max}-G_\text{min}\gg1$) in one or both dimensions. When designing the study, we we included only the trajectories with all differences and ranges in the lower 98th percentile, the total of 1,761 trajectories\footnote{The ``rough'' trajectories need to be analyzed differently.} and excluded the rest. Under the assumption that the observed processes were stationary, we translated each remaining trajectory $T$ to the origin by subtracting $\{t^U_1,P^U_1,F^U_1\}$ from each tuple in $T^U\!\in T$.

The trajectories retained in the data set demonstrate a remarkable positive correlation between blogging and socializing: active bloggers tend to acquire new friends at a faster rate than silent network members (Figure~\ref{Figure1}). This means that publishing and social activities in LiveJournal, when present, are highly synchronized. The details of this synchronization will be discussed in the next two sections.
 
\section{Microscopic Analysis\label{micro}}

At the microscopic level, we treat individual trajectories as timed sequences of events $e\in E=\{\Pi^+,\Phi^+,\Phi^-,$ $\Pi\Phi\}$, where symbols $\Pi$, $\Phi$, and $\Pi\Phi$ represent the changes in the number of posts, friends, and posts and friends together, and the sign in the superscript represents the direction of changes (but not the magnitude). Since 97\% of $\Pi$ events constitute an addition of $\le2$ posts and 97\% of $\Phi$ events constitute an addition or removal of
$\le7$ friends, we choose to see these events as binary---either occurring or not occurring---because the extent of variation is not significant given how minuscule it is relative to the time scope considered. The compound events $\Pi\Phi=\{\Pi^\pm\Phi^\pm\}$ are rare. We treat them as non-directional.

The delays between consecutive events for all trajectories are distributed exponentially with the
average rates of $0.45\,\text{day}^{-1}$ for publishing events and $0.2\,\text{day}^{-1}$ for social events.

For each trajectory $T$ and for each pair of event types $\{e_i,e_j\}\in E\times E$, we calculate $\psi^T_{ij}=\psi^T(e_j|e_i)$, the probability of an event $e_j$ immediately following
event $e_i$ along a certain trajectory $T^U$; in other words, 
the conditional probability of $e_j$ given $e_i$. Thus, $\psi^T=\{\psi^T_{\Pi^+\Pi^+},\psi^T_{\Pi^+\Pi^-},\ldots,\psi^T_{\Pi\Phi\,\Pi\Phi}\}$ is a vector in a 16-dimen\-sional metric space $\Psi$. We call this vector the signature of the trajectory. The proximity
of trajectories $T_1,T_2\in \Psi$, defined as the Euclidean distance $\Delta T=\|T_1-T_2\|$ between their signatures, corresponds to the similarity of the BSN
users' social and publishing behaviors.

\begin{figure}[tb!]\centering
\strut\epsfig{file=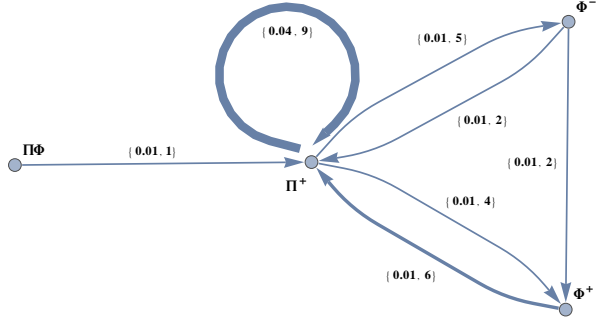,width=\columnwidth}
\caption{\label{Figure2}Micro cluster $\mu_7$ as a Markov chain. The edge labels show transition frequencies (in days$^{-1}$) and transition durations (in days).}
\end{figure}

\begin{table}[b!]
\begin{tabular}{rrll}
\hline  {\bf Id} & {\bf S} & {\bf Dominant Transitions} & {\bf Archetype}\\
\hline  
$\mu_1$& 398 & none & ``Reader'' \\ 
$\mu_2$  & 191 & none &  ``Reader''\\ 
$\mu_3$  &  187& \small{$\Pi^+\Pi^+$/3, $\Pi^+\,\Pi\Phi$/3, $\Pi\Phi\,\Pi^+\!/2$} &  ``B.-Soc.''\\ 
$\mu_4$  &  183& $\Phi^+\Phi^+$/10 &  ``Socializer''\\ 
$\mu_5$  &  158& none &  ``Reader''\\ 
$\mu_6$  &  148& $\Pi^+\Pi^+$/5 &  ``Blogger''\\ 
$\mu_7$  &  123& $\Pi^+\Pi^+$/9 &  ``Blogger''\\ 
$\mu_8$  &  110& \small{$\Phi^+\Phi^-$/4, $\Phi^-\Phi^-$/8, $\Phi^-\Phi^+\!/7$} &  ``Socializer''\\ 
$\mu_9$  &  99&  \small{$\Phi^+\Pi^+$/3, $\Pi^+\Phi^+$/3, $\Pi^+\Pi^+\!/3$}&  ``B.-Soc.''\\ 
$\mu_{10}$  &  90&  \small{$\Pi^+\Pi^+\!/2$, $\Pi^+\,\Pi\Phi/1$, $\Pi\Phi\,\Pi^+\!/3$}&  ``B.-Soc.''\\ 
$\mu_{11}$ &  78& none & ``Reader'' \\ 
$\mu_{12}$ &  74& $\Phi^+\Phi^+$/5 &  ``Socializer''\\ 
\hline 
\end{tabular} 
\caption{\label{Table1}Trajectory clusters, their sizes, dominant transitions (with
  transition durations, in days), and archetypes.}
\end{table}

We use the distance $\Delta T$ to group the trajectories and the corresponding BSN users into twelve
disjoint micro clusters $\mu_k=\{T\}$ with 74 to 398 trajectories per cluster (Table~\ref{Table1}). Each directory belongs exactly to one cluster. The number of micro clusters was chosen to match the number of the most significant macro clusters described in the next section. In our case,
this clustering method gives acceptably good results and is substantially simpler than similar methods proposed, e.g., in~\cite{Andrienko09,Moody11}. For each cluster $\mu_k$, we calculate its mean trajectory $T^{(k)}$ with the signature $\psi^{(k)}=\overline{\{\psi^T|T\in\mu_k\}}$.

To understand the collective behavior of cluster members, we model each cluster
as a Markov chain with social and publishing events representing
states, and pairs of consecutive events representing
transitions. Thus, e.g., events $\Phi^-$ and $\Pi^+$ immediately following one another
represent a transition between the states $\Phi^-$ and $\Pi^+$. The
probability of this transition in cluster $\mu_k$ is $\psi^{(k)}_{\Pi^+\Phi^-}$.

Based on $\psi^{(k)}$, we identified four user arche\-types
associated with each cluster (Table~\ref{Table1}): ``mainly bloggers''
(frequent publishing events), ``mainly socializers'' (frequent social
events), ``bloggers-socia\-lizers'' (both publishing and social events),
and ``readers'' (no events; the passive network members create
blogging accounts simply to read other people's blogs or leave comments).

As an example, consider clusters $\mu_{10}$, $\mu_3$, $\mu_6$, and
$\mu_7$. They have 90, 187, 148, and 123 trajectories, respectively,
with the dominant transition $\Pi^+\Pi^+$, ``(add a post) followed by
another (add a post)''. For the different clusters, this transition takes 2,
3, 5 or 9 days. These models correspond to more or less rigorous
``mainly bloggers.''

The Markov chain for the cluster $\mu_7$ is shown in
Figure~\ref{Figure2}. Transitions between the states $\Phi^+$,
$\Phi^-$, and $\Pi\Phi$ in the cluster are very rare: they happen
with the frequency of 0.01 days$^{-1}$. The self-loop around $\Pi^+$ is
more frequent, which is reflected by the thickness of the arc.

According to the microscopic analysis, 45\% of all users included in the study
are ``readers,'' 20\% are ``blog\-gers-socializers,'' 20\% are
``mainly socializers,'' and 15\% are ``mainly bloggers.''

\section{Macroscopic Analysis\label{macro}}
\begin{figure}[tb!]\centering
\strut\epsfig{file=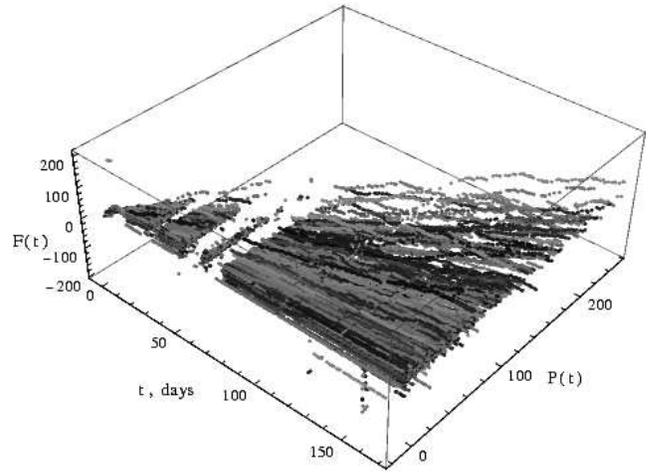,width=\columnwidth}
\caption{\label{Figure3}All user trajectories, translated to the origin. The gap around day 47 has been caused by networking problems.}
\end{figure}
At the macroscopic level, a unit of analysis is a blog trajectory
$\left\{t,P(t),F(t)\right\}$ in the time-friendship-publi\-shing space,
representing the user's social and blogging activities over
time (Figure~\ref{Figure3}). The translated trajectories radiate
from the origin approximately in the same direction. This is not very
surprising, given that friends are rarely unfriended~\cite{ahn2007} and posts, once published, are rarely deleted. However, the angle between the most extreme trajectories
is still large. We will use clustering again to identify similarly behaving
users.

The trajectories in Figure~\ref{Figure3} are remarkably smooth. For each component of each trajectory, we calculate the best-fit quadratic approximation $G(t)=a_0+a_1t+a_2t^2$. If $|a_2/a_1|<0.0085$, then we replace the quadratic model with a linear model $G(t)=a_0+a_1t$. Otherwise, a true quadratic model is used. Whether the trajectory bends in the direction away from the X axis or toward it, depends on the sign of $a_2/a_1$. If the best-fit linear approximation of $G(t)$ is unacceptably inaccurate with $R^2\!<\!0.7$, we treat $G(t)$ as constant in time.

\begin{figure}[tb!]\centering
\strut\epsfig{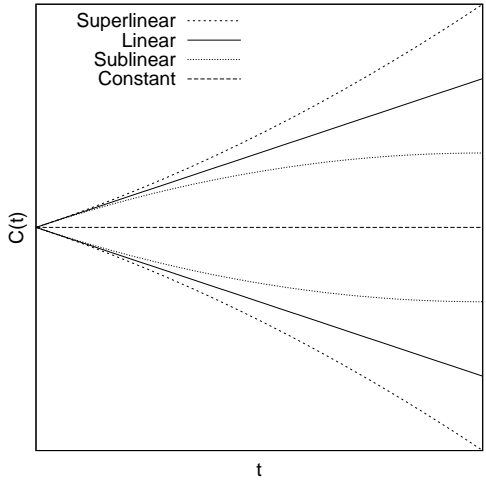}
\caption{\label{Figure4}Classification of trajectory dynamics.}
\end{figure}

Therefore, any $G(t)$ can be approximated using one of the following seven smooth functions (Figure~\ref{Figure4}):

\begin{itemize}
\item linear ($G''\!\approx\!0)$:
\begin{itemize}
\item ascending ($\uparrow: G'\!>\!0$),
\item constant ($\updownarrow: G'\!\approx\!0$) or
\item descending ($\downarrow: G'\!<\!0$),
\end{itemize}
\item quadratic locally ascending ($G'\!>\!0)$ bending up ($\upuparrows: G''\!>\!0$, superlinear) or down ($\upharpoonright: G''\!<\!0$, sublinear), or
\item quadratic locally descending ($G'\!<\!0)$ bending up ($\downharpoonright: G''\!>\!0$, sublinear) or down ($\downdownarrows: G''\!<\!0$, superlinear).
\end{itemize}

\begin{figure}[tb!]\centering
\strut\epsfig{file=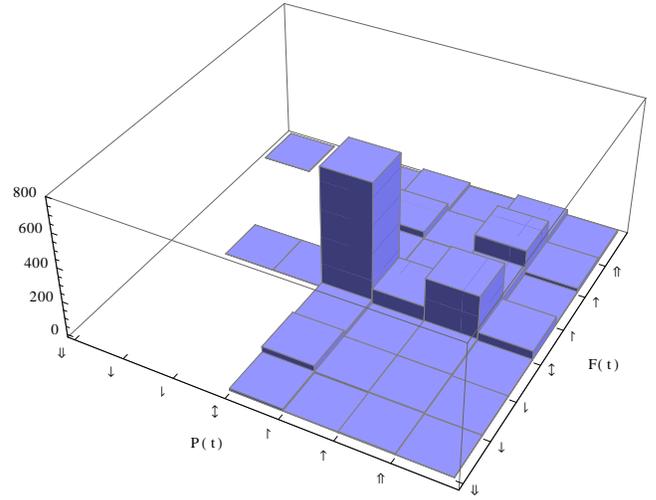,width=\columnwidth}
\caption{\label{Figure5}Joint distribution of $P(t)$ and $F(t)$ dynamics.}
\end{figure}

Overall, there are the total of seven possible behaviors for each of $P$ and
$F$, as described above. Each trajectory can be assigned to one of the $7\times7=49$ macro
clusters $M_{PF}$, based on the publishing dynamics $P$ and social
dynamics $F$ (Figure~\ref{Figure5}).

 The largest macro clusters are $M_{\updownarrow\updownarrow}$
 (``readers,'' both $P$ and $F$ are constant, 41\%), $M_{\uparrow\updownarrow}$ (``mainly
 bloggers,'' 15\%), and $M_{\uparrow\uparrow}$
 (``bloggers-socializers,'' 9\%). 

We assume that a trajectory belongs to a ``mainly blogger'' or to a
``mainly socializer'' if it has exactly one linear or superlinear $P$ or $F$
component, respectfully. If both components are linear or superlinear, then the
trajectory represents a ``blogger-socializer.'' Otherwise, it is a
``reader'' trajectory.

The clusters with anticorrelated publishing and social dynamics (e.g.,
growing number of posts vs shrinking number of friends) have
negligibly small membership (3\%).  LiveJournal posters are motivated to gain more friends, produce more posts, and increase their presence online.  Losing friends as a result of new posts may be a deterrent from writing more controversial posts.  Similarly, no one will gain friends as a result of deleting posts and reducing their presence online. In other words, social activity in
the selected subset of LiveJournal does not happen at the expense of publishing or the other
way around.

\begin{figure}[b!]\centering
\strut\epsfig{file=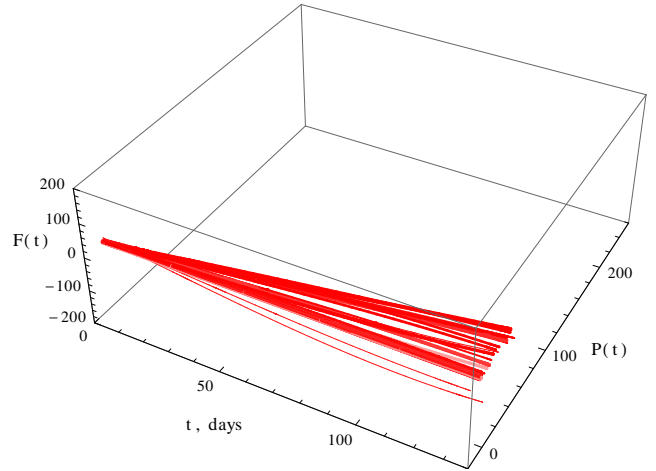,width=\columnwidth}
\caption{\label{Figure6}Mean trajectories for each macro cluster $M$. Line diameters are proportional to the $\log$ of the cluster size. Line darkness is proportional to the geometric mean fit quality $\sqrt{R^2_PR^2_F}$; cf. Figure~\ref{Figure3}.}
\end{figure}

Just as with the micro clusters, we calculated the mean trajectory for each macro cluster (Figure~\ref{Figure6}). We observed that the numbers of friends and posts along the mean trajectories in each cluster are loosely correlated: 
\begin{equation}
F(t)\approx9\sqrt{P(t)}.
\end{equation}

This relationship holds over the entire period of observation with $R^2\!\approx\!0.85$. We consider it as evidence to
the mainly blogging nature of LiveJournal, with the number of posts
dominating the number of friends most of the time.

\section{Comparison\label{compare}}

\begin{figure}[tb!]\centering
\strut\epsfig{file=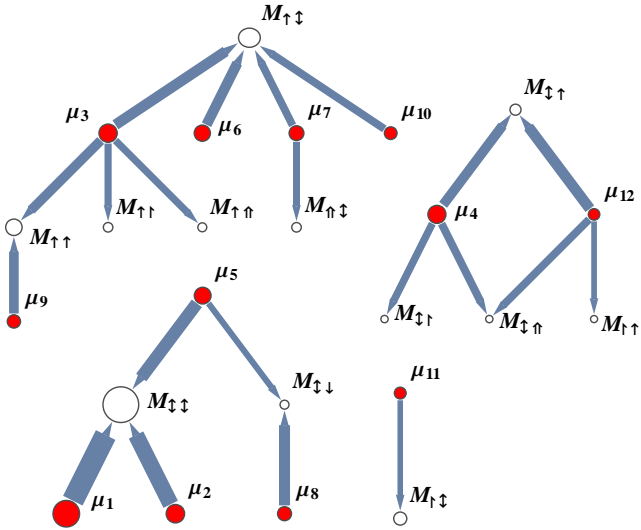,width=\columnwidth}
\caption{\label{Figure7}The correspondence between micro clusters $\mu_i$ and macro clusters $M_i$. The correspondence diagrams represent ``bloggers[-socializers]'' (top left); ``socializers'' (top right); ``readers'' (bottom). Edge thickness is proportional to the number of common elements. Vertex areas are proportional to the cluster sizes.}
\end{figure}

\begin{table}[b!]
\begin{tabular}{lrrr}
\hline  {\bf Archetype} & {\bf Micro} & {\bf Macro}&{\bf Difference}\\
\hline  
``Reader''& 45\% & 52\%&+7\%\\
\small{``Blogger-Socializer''}& 20\% & 15\%&-5\%\\
``Socializer''& 20\% & 11\%&-9\%\\
``Blogger''& 15\% & 22\%&+7\%\\
\hline
\end{tabular}
\caption{\label{Table2}Comparison of user archetypes obtained
through micro- and macroscopic analyses. }
\end{table}

Table~\ref{Table2} shows the comparison between the user
arche\-types obtained through micro- and macroscopic analyses. While
both methods confirm the dominance of the ``readers'' over the other
three archetypes, one can see that the macroscopic method consistently
underestimates the socializing component. This discrepancy is due to the
averaging nature of the macroscopic procedure. A trajectory of a
typical ``socializer'' may have a number of consecutive social events
$\Phi^+$ and $\Phi^-$, which will not change her number
of friends at the end of the observation period and thus will not be
detected by the macroscopic algorithm; such ``socializers'' will be
mistakenly assigned to the ``readers'' or ``bloggers'' categories. In other words, if many friend events occur, yet the net change is small, the macro analysis overlooks it.

The correspondence between the micro- and macroscopic clusters is
presented in Figure~\ref{Figure7}. The amount of overlap
between any two clusters is used as a measure of their similarity. The twelve
microscopic clusters and the twelve largest macroscopic clusters have
only 21 significant relationships (of those, one is one-to-one and
eleven are either many-to-one or one-to-many). This is just marginally more
than the minimum of twelve one-to-one relationships but substantially
less than the possible maximum of 12$\times$12=144 random relationships. 

Most micro clusters are connected to the macro clusters that represent
the same archetype. The only exceptions are the strong connections
between $M_{\uparrow\updownarrow}$ (``mainly bloggers'') and $\mu_3$
and $\mu_{10}$ (both ``bloggers-socializers'') and a weak connection
between $M_{\updownarrow\uparrow}$ (``mainly socializers'') and
$\mu_5$ (``readers''). The first exception confirms our hypothesis
about the macro method being more socially agnostic. At the moment, we
do not have an explanation of the second exception.

The sparsity of the resulting bipartite graph and its cohesion with
respect to the user archetypes implies that both clustering methods
describe the same taxonomy, but emphasize different nuances of network
dynamics.

\section{Conclusion\label{conclusion}}

We presented a study of joint social and publishing dynamics in
LiveJournal---a popular blogging social network (BSN). Over eighteen
hundred user accounts have been analyzed at both microscopic level
(represented as timed sequences of social and publishing events) and
macroscopic level (represented as trajectories in
temporal-social-publishing space). 

We have observed a significant general positive correlation between
blogging and socializing in LiveJournal. We also identified a number of user
archetypes that correspond to ``mainly bloggers,'' ``mainly
socializers,'' ``bloggers-socializers,'' and ``readers'' (passive
network members who create blogging accounts simply to read other
people's blogs). The analysis has been performed both at the microlevel
(individual posts and changes in contact lists modeled as Markov
chains) and at the macrolevel (trajectories in the
time-friendship-publishing space). Both approaches produced consistent
results:
\begin{itemize}
\item the majority of the BSN users are passive readers,
\item publishing is the dominant active behavior in a BSN, and
\item social activities, when present, complement blogging, rather than compete with it.
\end{itemize}

\section*{Acknowledgment}
This research was supported by National Science Foundation Grant $\#$1004996 ``Research Experience for Undergraduates.'' It has been partially presented as a poster at NetSci-2012, Evanston IL.

The authors are grateful to Kesann Walrond-McClean (currently at Amazon.com) for data collection and interpretation, professors Dan Stefanescu and Honggang Zhang, and other Suffolk University Research Experience for Undergraduates (REU) students for fruitful discussions.
 
\bibliographystyle{acm}
\bibliography{cs}
\end{document}